\documentstyle[preprint,pre,aps,eqsecnum]{revtex}
\tightenlines
\draft

\begin{document}

\title{Viscous fingering in liquid crystals:
Anisotropy and morphological transitions.}

\author{
R. Folch, J. Casademunt and A. Hern\'andez--Machado}
\address{
Departament d'Estructura i Constituents de la Mat\`eria\\
Universitat de Barcelona,
Avinguda Diagonal, 647, E-08028-Barcelona, Spain
}

\maketitle

\begin{abstract}

We show that a minimal model for viscous fingering with a nematic liquid
crystal in which anisotropy is considered to enter through two different
viscosities in two perpendicular directions can be mapped to
a two-fold anisotropy in the surface tension.
We numerically integrate the dynamics of the resulting problem with the 
phase-field approach to find and characterize
a transition between tip-splitting and side-branching as a function
of both anisotropy and dimensionless surface tension. This anisotropy
dependence could explain the
experimentally observed (reentrant) transition as temperature and applied
pressure are varied. Our observations are also consistent with
previous experimental evidence in viscous fingering within an etched cell and
simulations of solidification.

\end{abstract}


\newpage

\section{Introduction}

Interfacial instabilities arise in a wide variety of contexts, often of applied
interest, such as dendritic growth, directional solidification, flows in porous
media, flame propagation, electrodeposition, or bacterial growth 
\cite{general}.
Notwithstanding this disparity, 
there has been a search for unifying common features concerning the more
fundamental problem of their underlying nonequilibrium dynamics.
One such feature seems to be the role of anisotropy in determining the observed
morphology.

Thus, the finding that anisotropy is necessary for the needle
crystal to solve the steady state solidification problem 
(see e.g. \cite{brener}) and indeed a critical
amount of it 
to stabilize its tip and (possibly) generate side-branches in related local 
models \cite{locals}
motivated the inclusion of
anisotropy in viscous fingering experiments, either by engraving the plates
\cite{benjacob},
or by using a liquid crystal \cite{buka1,buka2,buka,canal}. 
In turn, these experiments in anisotropic 
viscous fingering confirmed the existence of a tip-splitting / side-branching
transition controlled by anisotropy 
and
driving force. 
Again, theoretical work on the solidification problem \cite{heine1} and 
numerical integration of its 
(nonlocal) dynamics \cite{heine2} showed the anisotropy in
the surface tension to control the transition in this system 
together with the dimensionless undercooling.
To our knowledge, no analytical or numerical work on anisotropic viscous 
fingering \cite{anisvf}
has focused on this transition so far. 

In this way, a picture that some kind of anisotropy can control the
tip-splitting or side-branching behavior of different systems emerged.
However, it is still not clear how different these anisotropies and systems
can be. 
On the one hand, not all kinds of anisotropy seem to control the
transition. The channel walls in a viscous 
fingering experiment, for instance, are known to play the same role as
surface tension anisotropy in free dendritic growth as far as the existence of
a single finger steady state solution with surface tension is concerned.
Moreover, even with isotropic
surface tension, this steady finger is stable up to a certain critical
amount of noise 
\cite{bensimon}, but no side-branching is observed, in contrast to free
dendritic growth.  
Above this threshold 
the tip splits, 
but again no side-branching is observed, in
spite of the anisotropy due to the channel walls. It is necessary to 
introduce some other type of anisotropy to observe the transition to 
side-branching.
On the other hand, some types of anisotropy not directly acting on the surface
tension \cite{ricard} seem to actually control the transition in some systems,
as is the case of liquid crystal viscous fingering experiments, which varied 
mainly the anisotropy in the viscosity, as well as etched cells, where the 
exact effect of the grooves on the free
boundary equations is unclear.
Therefore a connection between surface
tension anisotropy (seen to control the transition in the solidification
problem both experimentally and in simulations) 
and other types of anisotropy clearly lacks.
Here we present such a connection for the case of a simple model of a 
liquid crystal.

Specifically, we show that two different viscosities in two perpendicular
directions (in addition to some already anisotropic surface tension)
can be mapped to a two-fold surface tension anisotropy (times the rescaled
original anisotropic surface tension) through a convenient axis rescaling.
Moreover, we integrate the resulting problem to confirm the existence of the
morphological transition also for the viscous fingering equations.
The numerics use a previously developed \cite{pf1} and 
thoroughly tested \cite{pf2} phase-field
model for viscous fingering.

We do find such a transition as a function of the amount of anisotropy
and the value of the dimensionless surface tension itself (i.e., 
the driving force for fixed surface tension).
The results are consistent with experimental results in viscous fingering
with a liquid crystal \cite{buka1,buka2,buka,canal}
and etched cells \cite{benjacob}, and
also with theory and simulations for the solidification problem 
\cite{heine1,heine2}.

The layout of the rest of the
paper is as follows: in Sec. \ref{secmod} we refer to the special features
of liquid crystals concerning viscous fingering and present a simple model
for them.
We then map this model onto the basic Saffman-Taylor
problem with a two-fold anisotropy in the surface tension. In Sec. \ref{secpf}
we briefly describe the phase-field model used and present the numerical
results. Finally, in Sec. \ref{secdis} we discuss their consequences for
viscous fingering with a liquid crystal and consistency with related problems.
 
\section{Model}
\label{secmod}

In the nematic phase of a liquid crystal its molecules are locally
oriented, giving rise to anisotropy in the viscosity and surface tension. The
degree of orientation depends on the proximity of the other phase(s), namely
the isotropic (and for some liquid crystals the smectic), i.e., it still
depends on temperature, and so does the anisotropy, mainly in the viscosity
\cite{buka}.
Therefore one should be able to explain the tip-splitting / side-branching
transition as a function of temperature in the nematics by means of the 
anisotropy in the viscosity alone.

In a viscous fingering experiment, the director forms a small angle with the
velocity field, except maybe for the neighborhood of the interface, where it
might follow its normal direction. So, as a first approximation, 
one can consider that there is flow alignment, and, therefore, a
velocity dependent viscosity, which would make the flow nonlaplacian. However,
in the vicinity of a finger tip we can approximate the direction of the flow 
for that of the finger, so that we can make a minimal model with only two 
different viscosities: one in 
the direction parallel to the finger and one in the perpendicular one.
More details can be found in Ref. \cite{buka}.

Let us now review the formulation of the Saffman--Taylor equations to account
for those two different viscosities in two perpendicular directions
$x$ and $y$. We will do it for the channel geometry, although the result also
applies to the circular cell used in the experiments of Ref. \cite{buka}
with minor changes. (i.e. two different viscosities in the radial and
tangential directions would also map to the standard viscous fingering
equations and the same functional dependence for the surface tension
anisotropy).
For the sake of generality, we consider both the
displaced ($1$) and the injected ($2$) fluid to have a certain distinct
viscosity ($\mu_1$,$\mu_2$).

As in the usual Saffman--Taylor problem, in each bulk we assume the flow to 
be incompressible,
\begin{equation}
\label{eq:incompressibility}
\vec \nabla\cdot\vec u=0
\end{equation}
(where $\vec u$ is the fluid velocity in the reference frame moving
 with the mean interface at $V_\infty$, the injection velocity),
and also Darcy's law to hold, but now for two different viscosities 
$\mu_{x,i}$,
$\mu_{y,i}$,
\begin{eqnarray}
\label{eq:Darcy}
u_x&=&-\frac{1}{\eta_{x,i}}\partial_x p\\
\nonumber
u_y&=&-\frac{1}{\eta_{y,i}}\left(\partial_y p +\rho_i g_{eff}
\right)-V_\infty,
\end{eqnarray}
where $i=1,2$ stand for each fluid, $u_x$, $u_y$ are the $x$, $y$ components 
of $\vec u$, $p$ is the
pressure, $\eta_{x,i}=(12/b^2)\mu_{x,i}$ is an inverse mobility in the
$x$ direction, $\eta_{y,i}=(12/b^2)\mu_{y,i}$, in the $y$ direction, 
$\rho_i$, the density, and
$g_{eff}$, the effective gravity in the plane of the channel.
Also as in the usual Saffman--Taylor problem, on the interface the normal
velocity is continuous and equals that of the interface,
\begin{equation}
\label{eq:vn}
\hat r\cdot\vec u_1=\hat r\cdot\vec u_2=v_n
\end{equation}
(where $r$ is a coordinate perpendicular to it increasing towards fluid $1$
 and $v_n$ its normal velocity),
and the pressure has a jump given by Laplace' law,
\begin{equation}
\label{eq:pressure}
p_1-p_2=\sigma(\phi)\kappa,
\end{equation}
with $\sigma(\phi)$ the (anisotropic) surface tension and $\kappa$ the 
interface curvature.
 
Due to Eqs. (\ref{eq:incompressibility}), and (\ref{eq:vn}), the flow can be 
described by a 
scalar field
$\psi$, the stream function, defined even on the interface by 
$u_x=\partial_y \psi$, $ u_y=-\partial_x \psi$ (see e.g. Refs. 
\cite{pf1,aref}). However, because of the different viscosities in the
$x$ and $y$ directions, the problem is nonlaplacian (there is vorticity) 
in the bulk:
\begin{equation}
\label{eq:nonlaplacian}
\nabla^2\psi=
-|\vec\nabla\times\vec u|\neq 0.
\end{equation}

To circumvent this, we rescale the $x$ and $y$ axis by a different factor. We 
also take advantage to adimensionalize the resulting equations in the same way 
as in Refs. \cite{pf1,aref}, so that they can be compared to those in these
references, and especially to Ref. \cite{pf1} in order to generalize the
phase-field model described there to the case of anisotropic viscosity. Thus,
we perform the following change of variables:
\begin{eqnarray}
\nonumber
x=a_x\tilde{x},\\
\label{eq:change}
y=a_y\tilde{y},\\
\nonumber
t=\frac{W}{U_*}\tilde{t},
\end{eqnarray}
where tildes
denote new variables, $a_x,a_y$ have units of length,
$U_*$ is a velocity, and $W$, the channel width. We find
\begin{equation}
\label{eq:incompressibility2}
\vec \nabla\cdot\vec u=\frac{U_*}{W}\tilde{\vec \nabla}\cdot\tilde{\vec u}=0,
\end{equation}
so that the flow is still incompressible and we can define a new stream
function $\tilde{\psi}=(W/U_*)\left[\psi/(a_x a_y)\right]$, which will be 
laplacian in the bulk if and only if
[see Eq. (\ref{eq:nonlaplacian})] the velocity field is potential in each
fluid,
\begin{equation}
\label{eq:potential}
\tilde{\vec u}=-\frac{W}{U_*}\left[\frac{1}{W^2\tilde{\eta}_i}
\left(\tilde{\vec\nabla} p + a_y\rho_i g \hat y\right)
+\frac{V_\infty}{a_y}\hat y\right]
\end{equation}
which is now the case as long as we choose $a_x,a_y$ to be such that
\begin{equation}
\label{eq:choice}
a_x^2\eta_{x,i}=a_y^2\eta_{y,i}\equiv W^2\tilde{\eta}_i.
\end{equation}

On the interface, Eq. (\ref{eq:vn}) will be formally unchanged as
long as the choice of $a_x,a_y$ is the same at both sides. Note that,
according to Eq. (\ref{eq:choice}), this
implies that the ratio $m\equiv\eta_x/\eta_y$ must be the same for both
fluids. In an air-liquid crystal experiment, this is obviously not the case, 
but, in the limit in which the viscosity of the air is negligible
compared to that of the liquid crystal, the anisotropic character of the air
viscosity in our model becomes irrelevant. In terms of the stream function,
Eq. (\ref{eq:vn}) for the new variables then reads
\begin{equation}
\label{eq:continuity}
\partial_{\tilde{s}} \tilde{\psi}_1=
\partial_{\tilde{s}} \tilde{\psi}_2=-\tilde{v}_n,
\end{equation}
where $s$ is the arclength along the interface and such that
$\hat s \times \hat r =
\hat x \times \hat y$. 

As for $\partial_{\tilde{r}}\tilde{\psi}$, the boundary condition for it will
be given by that for
$\tilde{u}_{\tilde{s}} \equiv \tilde{\hat s}\cdot \tilde{\vec u}$. Indeed, it
will have a jump on the interface due to the fact that $\tilde{\vec u}$ is not
potential on the very interface [see Eq. (\ref{eq:potential}]
because of the jump in $\tilde{\eta}_i$, which
gives rise to a singular vorticity on it:
\begin{eqnarray}
\nonumber
\;&\;&\;\frac{(\tilde{\eta}_1+\tilde{\eta_2})
(\tilde{u}_{\tilde{s},1}-\tilde{u}_{\tilde{s},2}) + 
(\tilde{\eta}_1-\tilde{\eta_2})
(\tilde{u}_{\tilde{s},1}+\tilde{u}_{\tilde{s},2}) }{2}\\
\label{eq:jump}
&=&
\tilde{\eta}_1\tilde{u}_{\tilde{s},1}-\tilde{\eta_2}\tilde{u}_{\tilde{s},2}\\
\nonumber
&=& -\frac{W}{U_*} \left\{
\frac{1}{W^2}\partial_{\tilde{s}}(p_1-p_2) +
\left[\frac{a_y}{W^2}g(\rho_1-\rho_2) +
\frac{(\tilde{\eta}_1-\tilde{\eta_2})V_\infty}{a_y} \right]
\hat y\cdot\tilde{\hat s} \right\}
\end{eqnarray}
and therefore, making use of Eq. (\ref{eq:pressure}),
\begin{eqnarray}
\nonumber
\;&\;&\; \partial_{\tilde{r}}\tilde{\psi}_1-
\partial_{\tilde{r}}\tilde{\psi}_2 =
\tilde{u}_{\tilde{s},1}-\tilde{u}_{\tilde{s},2} \\
\label{eq:discontinuity}
&=& -\frac{2}{U_*} \left\{
\frac{1}{W^2(\tilde{\eta}_1+\tilde{\eta_2})}
\partial_{\tilde{s}}[\sigma(\phi)W\kappa] +
\left[\frac{a_y}{W}
\frac{g(\rho_1-\rho_2)}{(\tilde{\eta}_1+\tilde{\eta_2})}
+\frac{W}{a_y}cV_\infty \right]
\hat y\cdot\tilde{\hat s} \right\}\\
\nonumber
\;&\;&\; -c\left(\partial_{\tilde{r}}\tilde{\psi}_1 +
\partial_{\tilde{r}}\tilde{\psi}_2 \right),
\end{eqnarray}
where $c\equiv(\tilde{\eta}_1-\tilde{\eta_2})
/(\tilde{\eta}_1+\tilde{\eta_2})$. Now choosing 
$a_y/a^2_x=1/W$,
Eq. (\ref{eq:choice}) yields $m=a_y/W$, and defining
$U_*\equiv cV_\infty/m+\left[mg(\rho_1-\rho_2)\right]
/(\tilde{\eta_1}+\tilde{\eta_2})$ we recover the usual result for viscous
fingering in a channel (see Refs. \cite{pf1,aref}),
\begin{equation}
\label{eq:discontinuity2}
\partial_{\tilde{r}}\tilde{\psi}_1-
\partial_{\tilde{r}}\tilde{\psi}_2 =
-2\partial_{\tilde{s}}[B(\phi)W\kappa]
-2\hat y\cdot\tilde{\hat s}
-c\left(\partial_{\tilde{r}}\tilde{\psi}_1 +
\partial_{\tilde{r}}\tilde{\psi}_2 \right),
\end{equation}
with $B(\phi)\equiv 
\sigma(\phi)/\left[W^2(\tilde{\eta}_1+\tilde{\eta_2})U_*\right]$, except for
the $m$ factors in the definition of $U_*$ ---and therefore in $B(\phi)$---
and the clue fact that $W\kappa$ and $\sigma(\phi)$ are still 
in the old variables and must be rescaled:
\begin{eqnarray}
\label{eq:curvature}
\;&\;&\; \kappa\equiv \frac{d^2y}{dx^2}
\left[1+\left(\frac{dy}{dx}\right)^2\right]^{-3/2}\\
\nonumber
&=& \frac{a_y}{a_x^2}\frac{d^2\tilde{y}}{d\tilde{x}^2}
\left[1+\left(\frac{a_y}{a_x}
\frac{d\tilde{y}}{d\tilde{x}}\right)^2\right]^{-3/2}
=\frac{1}{W}\frac{d^2\tilde{y}}{d\tilde{x}^2}
\left[1+m\left(\frac{d\tilde{y}}{d\tilde{x}}\right)^2\right]^{-3/2}, 
\end{eqnarray}
so that we
obtain
\begin{equation}
\label{eq:wcurvature}
W\kappa=\tilde{\kappa}\left[\frac{1+\left(d\tilde{y}/d\tilde{x}\right)^2}
{1+m\left(d\tilde{y}/d\tilde{x}\right)^2}\right]^{3/2}
=\frac{\tilde{\kappa}}{\left[1+(m-1)\cos^2\tilde{\phi}\right]^{3/2}},
\end{equation}
where $\tilde{\phi}$ is the angle from $\hat x$ to $\hat{\tilde{r}}$.

To summarize, we recover the usual viscous fingering equations, including
Eq. (\ref{eq:discontinuity2}), which finally reads
\begin{equation}
\label{eq:discontinuity3}
\partial_{\tilde{r}}\tilde{\psi}_1-
\partial_{\tilde{r}}\tilde{\psi}_2 =
-2\partial_{\tilde{s}}[\tilde{B}(\tilde{\phi})\tilde{\kappa}]
-2\hat y\cdot\tilde{\hat s}
-c\left(\partial_{\tilde{r}}\tilde{\psi}_1 +
\partial_{\tilde{r}}\tilde{\psi}_2 \right),
\end{equation}
but now with an anisotropic dimensionless surface tension of the form 
\begin{equation}
\label{eq:banis}
\tilde{B}(\tilde{\phi})=\tilde{B}_0 \times \tilde{\Sigma}(\tilde{\phi}) \times
\left[\frac{1}{1+(m-1)\cos^2\tilde{\phi}}\right]^{3/2},
\end{equation}
where $\tilde{B}_0$ is the dimensionless surface tension of isotropic viscous
fingering
\begin{equation}
\label{eq:b}
\tilde{B}_0\equiv \frac{\sigma_0}
{W^2\left[(\tilde{\eta}_1-\tilde{\eta_2})V_\infty/m
+mg(\rho_1-\rho_2)\right]}
\end{equation}
except for the $m$ factors,
with $\sigma(\phi)\equiv \sigma_0 \Sigma(\phi)$.
This means that, even with an originally isotropic surface tension, that in the
rescaled problem has a two-fold anisotropy with a very specific form
given by the last (third) factor at the r.h.s. of Eq. (\ref{eq:banis}). On the
other hand, the possible original anisotropy in the surface tension will
change its functional form in the rescaled problem according to
\begin{equation}
\label{eq:anisotropy}
\tilde{\Sigma}(\tilde{\phi})=\Sigma(\phi)=
\Sigma\left[\arctan (\sqrt{m}\tan\tilde{\phi})\right],
\end{equation}
and the rescaled problem will have the two-fold anisotropy of the mentioned
last factor superimposed to the transformed anisotropy of
Eq. (\ref{eq:anisotropy}) [second factor at the r.h.s. of 
Eq. (\ref{eq:banis})].
A similar result was found in a different context, namely for the 
nematic-smectic B transition, where two different heat diffusivities in two
perpendicular directions could be mapped to the same type of anisotropy in
the surface tension and the same type of transformation in the
original anisotropy \cite{kramer}.
However, note that here the assumption is that the growth is in the direction
of lowest viscosity (because of flow alignment of the director), which results
in growth in the direction of largest surface tension ($\tilde{\phi}=\pi/2$),
whereas in Ref. \cite{kramer} the situation is just the opposite:
growth was found to be in the direction of
lowest diffusivity because that is the direction of lowest capillary length.
Also for isotropic diffusivities it is known that steady needle crystals
can only grow in the direction of minimal
capillary length \cite{brener}, although there the anisotropy is 
assumed to be four-fold.

Finally, for the described minimal model the original anisotropy in the surface
tension would be two-fold, e.g.
\begin{equation}
\label{eq:2fold}
\Sigma(\phi)=1-\alpha\cos^2\left(\phi-\frac{\pi}{2}\right),
\end{equation}
so that the transformed anisotropy would read
\begin{equation}
\label{eq:2foldr}
\tilde{\Sigma}(\tilde{\phi})=1-\frac{m\alpha
\cos^2\left(\tilde{\phi}-\pi/2\right)}
{1+(m-1)\cos^2\tilde{\phi}}.
\end{equation}

\section{Numerical Integration}
\label{secpf}

We now integrate the rescaled problem, namely 
Laplace equation for the stream function with the boundary
conditions Eqs. (\ref{eq:continuity}) and (\ref{eq:discontinuity3}).
In principle, given an initial condition, we should rescale it, evolve it
using the rescaled dynamics, and translate the resulting interface back to
the original variables, but we will not perform any rescaling, since the
initial condition is free, and the tip-splitting or side-branching character
of the result is unaffected by the final
translation into the original variables. Instead, we will consider the 
rescaled problem in its own, and simulate it by means
of the following phase-field model,
\begin{equation}
\label{eq:sf}
\tilde{\epsilon} \frac{\partial\psi}{\partial t}=\nabla^2\psi+c\vec \nabla \cdot
(\theta \vec \nabla \psi)+\frac{1}{\epsilon} \frac{1}{2\sqrt 2}
\gamma(\theta )(1-\theta^2)
\end{equation}
\begin{equation}
\label{eq:pf}
\epsilon^2 \frac{\partial \theta}{\partial t}=f(\theta)+\epsilon^2\nabla^2\theta
+\epsilon^2 \kappa(\theta ) |\vec \nabla \theta |+\epsilon^2
\hat z \cdot
(\vec \nabla \psi \times \vec \nabla \theta),
\end{equation}
where $\theta$ is the phase field,
$\epsilon$, $\tilde{\epsilon}$ are model parameters which must be small to
recover the sharp-interface equations of the rescaled problem,  
and we have dropped the tildes of the rescaled variables.
We have defined
$f(\theta )\equiv \theta (1-\theta^2)$, and
$\frac{\gamma(\theta)}{2}\equiv \hat s(\theta)\cdot\left[\vec\nabla
B(\theta)\kappa(\theta)
+\hat y\right]$,
$\kappa(\theta)\equiv -\vec\nabla \cdot \hat r(\theta)$,
with
$\hat r(\theta)\equiv \frac{\vec\nabla \theta}{|\vec\nabla \theta|}$
and $\hat s(\theta)\equiv \hat r(\theta) \times \hat
z$.
This model was introduced for isotropic viscous fingering
in Ref. \cite{pf1}
and extensively tested in Ref. \cite{pf2}. 
From this work we know that it will yield converged results for the
steady 
fingers (and, in particular, for their
widths) if both $\epsilon\leq 0.2\sqrt{B_0}$ and 
$\tilde{\epsilon}\leq 0.2(1-c)$. 
The only change to be made for the
anisotropic case is to set $B(\theta)$ not merely equal to a constant, but
to that given by Eq. (\ref{eq:banis}) taking
$\phi=\phi(\theta)=\arccos \hat x\cdot\hat r(\theta)$. This gives 
$B(\theta)=B(\phi)+O(\epsilon^3)$, which not only satisfies the desired
sharp-interface limit, but also ensures that the introduction of anisotropy 
will not result in any extra first-order correction to the free boundary
problem, so that the above conditions
on $\epsilon,\tilde{\epsilon}$ still hold.

The same phase-field equations could be used for the circular geometry
reinterpreting the parameters, since an analogue rescaling yields formally
the same result. However, the boundary conditions would change. For instance,
injection at the center of the cell should also be considered.
Anyhow, we choose to simulate the well controlled
situation in which an (unstable) steady
finger propagates in a channel of width $W$. This representation is of
course exact
for single fingers in experiments carried out in the linear Hele--Shaw cell
\cite{canal}, but only a (good)
approximation for the vicinity of a finger in a multifinger configuration
(with many fingers) in
the circular geometry \cite{buka}.

We investigate the transition between the tip-splitting and side-branching
behaviors as both the dimensionless surface tension $B_0$ and its anisotropy
$m-1$ are varied. We are interested mainly in the effect of the anisotropy 
coming from the viscosity ($m-1$), so we drop that in the original surface 
tension ($\sigma(\phi)=\sigma_0$). The runs use equal viscosities in both 
fluids, $c=0$, for reasons of numerical efficiency ($\tilde{\epsilon}=0.2$), 
but we do not expect the viscosity contrast to affect the 
stability of the tip for similar reasons for which it does not play any
role in the (linear) stability of a flat interface. To check this conjecture
we ran simulations with $B_0=10^{-3}$ both for $c=0$ and $c=0.8$, two values
of the viscosity contrast for which a dramatic change in the competition 
dynamics was seen using the same phase-field model \cite{pf2}, and we found
that the transition lies in both cases between $m=2$ and $m=2.25$. These 
$c=0.8$ runs are indeed the ones shown in Figs. 1 and 2.

We use $\epsilon=4\times 10^{-3}$, so that we can simulate accurately with 
values of the dimensionless surface tension down to $B_0=4\times 10^{-4}$.
We first run a steady finger with a large, isotropic dimensionless surface
tension $B(\phi)=B_0=10^{-2}$, large enough for the finger not to destabilize
for the amount of (numerical) noise we have
and thus reach its steady width and velocity. Once this is achieved ---see
inner interface in Figs. 1(a) and 1(b)---, we perform a ``quench'' in surface
tension, i.e., we instantly reduce it to some lower value. Simultaneously,
we also introduce some amount of anisotropy $m-1$. 

The subsequent interface
evolution for
$B_0=10^{-3}$ (and $c=0.8$, $\tilde{\epsilon}=0.08$) is also shown 
within the reference frame moving with the mean interface
in Figs. 1(a) ($m=2$) and 1(b) ($m=2.25$) in the
form of snapshots at time
intervals $0.11$. (Simulations used
only half of the channel and reflecting boundary conditions at its center
$x=0$). The corresponding $y$ position of the interface 
at the center
of the channel 
(also in the frame of the mean interface)
is plotted against time in Fig. 2. 

For this value of $B_0$ the
finger clearly destabilizes: First its tip widens and flattens (see Fig. 1)
and therefore
slows down (see Fig. 2) for any value of the anisotropy. (Note that for $t<0$
the tip position would be a straight line in time, since the finger was steady,
and, in particular, its velocity).
Then, for $m=2$ the tip continues
to flatten and slow down until its curvature [Fig. 1(a)]
and eventually its velocity 
in the frame of the mean interface (lower curve
in Fig. 2) reverse their signs. Finally, the velocity of the interface 
at the center of the channel seems to reach some negative constant value
(again, in the frame moving with the mean interface)
corresponding to the growth of two parallel fingers at each side. We identify 
this reversal of the curvature sign at the center of the finger and this
always convex
tip position vs. time plot with the tip-splitting morphology. 

In contrast, for $m=2.25$ the reversal of the curvature sign takes place at
some distance of the center of the channel, while at the center the
curvature increases again [Fig. 1(b)]
and makes it possible for the tip to speed up again
as well, giving rise to a change of concavity in the tip position vs. time 
plot (upper curve in Fig. 2). We identify this reversal of the curvature sign
at a distance from the center of the channel and this change of concavity
in the tip position vs. time plot with the side-branching morphology.  

In this way we systematically explore values of the dimensionless surface 
tension $B_0$ ranging from $B=10^{-2}$ down to $B=4\times 10^{-4}$. For each
value of $B_0$ we simulate with several values of the anisotropy $m-1$,
and we find that there is a relatively sharp transition between 
the tip-splitting and 
side-branching morphologies. 
In Fig. 3 we show for each value of $B_0$ ($x$
axis) the two closest values of $m-1$ ($y$ axis) for which the two different 
morphologies are observed, namely tip-splitting (circles) and side-branching
(triangles). Thus we know that the transition line must lie somewhere between
the circles and the triangles, and that above (larger values of $m-1$) and
left (lower values of $B_0$) of that transition line the morphology is
side-branching, and below and right of it, tip-splitting.  
This means that the critical anisotropy $m-1$ above which side-branching
replaces tip-splitting decreases with decreasing dimensionless surface tension
$B_0$. 

In fact, this critical anisotropy vanishes at
$B_0\sim 5\times 10^{-4}$, and below this value only side-branching is
observed, even if one uses negative anisotropies down to $m-1=-0.9$, which
correspond to a viscosity larger in the direction of growth of the finger than
in the perpendicular one, and which is not the case of the liquid crystal
experiments that motivated this study. ($m-1>-1$ to keep the two viscosities
and therefore $B(\phi=0)$ finite and positive). 
Of course, the specific value of $B_0$ for which the critical anisotropy 
vanishes could be affected by the fact that a residual 
(four-fold) grid anisotropy remains, but it seems unavoidable that there is
such a (finite) value of $B_0$, since the transition line curves down as
$B_0$ is decreased, and for large enough values of $B_0$ or the anisotropy
$m-1$ the grid spacing
$\Delta x=\epsilon=4\times 10^{-3}$ is far too fine to affect the effective 
anisotropy.

On the other hand, for 
$B_0\geq 1.4\times 10^{-3}$ and for the time elapsed in our runs no
clear side-branching is actually observed above the transition line
extrapolated from lower values of $B_0$, whereas tip-splitting still occurs
below that line. For even larger values of $B_0$, namely 
$B_0\sim 2\times 10^{-3}$, not even tip-splitting is observed again within
the time elapsed, although the steady finger still destabilizes through the
widening and flattening of its tip. Finally, for $B\geq 10^{-2}$ the finger is
completely stable for the amount of noise we have, as was pointed out before.

\section{Discussion and conclusions}
\label{secdis}
We have shown that viscous fingering with two different viscosities in two 
perpendicular directions maps to the standard
viscous fingering equations (i.e., for isotropic viscosity)
with an extra two-fold anisotropy in the surface
tension, which is such that, together with the hypothesis of flow alignment of 
the director,
leads to growth in the direction of maximal surface tension.
We have simulated the resulting problem using a previously developed
phase-field model \cite{pf1,pf2}, and we have found that there is a transition
from the tip-splitting to the side-branching morphology as either the
anisotropy in
the surface tension is increased or the dimensionless surface
tension is decreased. We now draw the connection with the liquid crystal
experiments of Ref. \cite{buka}.

The observed anisotropy 
dependence is consistent with
the experimental finding that there is a transition from 
tip-splitting to side-branching and back to tip-splitting with temperature 
in the nematic
phase, since close to the other phases the director alignment, and consequently the
anisotropy, weaken \cite{buka}.
The transition is found to be also reentrant with injection
pressure \cite{buka}, 
which is explained there with the hypothesis that too low 
pressures do not achieve flow alignment, whereas too large ones break down
the Hele--Shaw approximation because of the importance of inertial terms in 
the hydrodynamic equations, which then destroy the flow alignment again.
This anisotropy dependence is also consistent with simulations of the
boundary layer model \cite{locals} and the full solidification problem
\cite{heine2}, as well as with analytical approaches to solidification
\cite{heine1}.

As for the dimensionless surface tension $B_0$ dependence, 
one first needs to relate
the values of $B_0$ used in the channel
simulations to the experimental parameters in the circular geometry. To do
this we consider a virtual channel whose walls are placed at half the distance
between a finger and its nearest neighbors. The channel width $W$ is given by
this distance between adjacent finger tips, whereas the effective injection
velocity $V_\infty$ turns out to be the ratio between the injection pressure 
and  $R$, the mean distance between a tip and the injection point. Then, the 
following dynamic picture of a typical experiment in the 
circular cell emerges:
Initially some fingers develop. If the
anisotropy, $m-1$, is strong enough, their tips are stable (which corresponds
to a point
above the transition line in Fig. 3, where the observed morphology is
side-branching). As these fingers grow radially,
$W$ increases as $R$, whereas the effective driving force,
$(\eta_1-\eta_2) V_\infty$, decreases as $1/R$, so that the dimensionless
surface tension $B_0$ they experience is found to decrease as $1/R$.
Thus, the corresponding point in Fig. 3 moves to the left 
(lower values of $B_0$), and the side-branching behavior is preserved.

In contrast, if the anisotropy $m-1$ is not strong enough, the tips split
(the corresponding point is below the transition line in Fig. 3, 
where the observed morphology is tip-splitting). 
As a result the number of fingers increases,
which then 
compensates for the growth of the distance between finger tips
as $R$ in such a way that
the effective dimensionless surface tension $B_0$ keeps roughly steady during 
the pattern development, so that the corresponding point in Fig. 3
basically does not move. Thus
the transition line is not crossed and the 
tip-splitting behavior is also preserved.
In this case, we can estimate $B_0$ in the experiments to be of order 
$10^{-3}$, whereas $m$ is said to
be around 2 at the transition \cite{buka}. This is indeed very close to our
transition point $B=10^{-3},2.05\leq m\leq 2.1$ in Fig. 3, 
but it should be taken into account that the value of
$B_0$ below which fingers destabilize is known to depend on the amount of noise
present \cite{bensimon}. Accordingly, one expects the whole
transition curve to be shifted to different values of $B_0$ for a different
amount of noise. The latter, however, is uncontrolled both in the experiments 
and the simulations.

On the other hand, 
the dependence of the value of the anisotropy at the transition
on the driving force has been seen in
viscous fingering with an etched cell \cite{benjacob}. However, we find
the critical amount of anisotropy
for side-branching to 
replace tip-splitting to vanish at a finite value of $B_0$. Below this value
we only observe side-branching.

A more realistic model of viscous fingering in a liquid crystal should include
a velocity dependent viscosity. In general the resulting nonlaplacian character
of the problem could not be avoided, but in principle it would still be
possible to simulate the dynamics by means of a phase-field model.

\section*{Acknowledgements}
We are grateful to \'A. Buka and T. T\'oth-Katona
for drawing our attention to this problem and for helpful discussions.
We acknowledge financial support from the Direcci\'on
General de Ense\~{n}anza Superior (Spain), Projects No. PB96-1001-C02-02 and
PB96-0378-C02-01,
and the European Commission Project No. ERB FMRX-CT96-0085.
Simulations have been carried out using the resources at CESCA and
CEPBA, coordinated by $\rm C^4$.
 R.F. also acknowledges
a grant from the Comissionat per a Universitats i Recerca
(Generalitat de Catalunya).

\newpage
\section*{figure captions}

\paragraph*{Fig. 1}
Destabilization of the tip of a (stationary) finger after instantly
decreasing $B_0$ to $B_0=10^{-3}$ at the time of the first
interface shown.
Successive interfaces are shown in the reference frame moving with the mean 
interface
at time intervals 0.11, for $c=0.8$, $\tilde{\epsilon}=0.08$. The latest
interface is represented in bold.
(a)Tip-splitting for $m=2$. (b) Side-branching for $m=2.25$.

\paragraph*{Fig. 2}
$y$ coordinate of the interface at the center of the channel (x=0)
in the reference frame of the mean interface as a function
of time corresponding to Figs. 1(a) (lower curve) and 1(b) (upper curve).
$t=0(0.88)$ corresponds to the first (last) interface shown there.

\paragraph*{Fig. 3}
Transition between tip-splitting (circles) and side-branching (triangles)
as a function of the surface tension anisotropy $m-1$ and the dimensionless
surface tension $B_0$.


\begin{references}

\bibitem{general}
J. S. Langer, Rev. Mod. Phys. {\bf 52}, 1 (1980);
D. A. Kessler, J. Koplik and H. Levine, Adv. Phys. {\bf 35}, 255 (1988);
P. Pelc\'{e}, in {\it Perspectives in Physics} (Academic, New York, 1988).

\bibitem{brener}
See E. A. Brener and V. I. Mel'nikov, 
Adv. Phys., {\bf 40}, 53 (1991) and references therein.

\bibitem{locals}
D. Kessler, J. Koplik and H. Levine, Phys. Rev. A {\bf 31}, 1712 (1985);
E. Ben-Jacob, N. D. Goldenfeld, B. G. Kotliar and J. S. Langer, Phys. Rev.
Lett. {\bf 53}, 2110 (1984);
E. Ben-Jacob, N. D. Goldenfeld, J. S. Langer and G. Sch\"on, Phys. Rev. Lett.
{\bf 51}, 1930 (1983); Phys. Rev. A {\bf 29}, 330 (1984);
R. Brower, D. Kessler, J. Koplik and H. Levine, Phys. Rev. Lett. {\bf 51},
1111 (1983); Phys. Rev. A {\bf 29}, 1335 (1984); D. Kessler, J. Koplik
and H. Levine, Phys. Rev. A {\bf 30}, 3161 (1984). 

\bibitem{benjacob}
E. Ben-Jacob, R. Godbey, N. D. Goldenfeld, J. Koplik, H. Levine, T. Mueller
and L. M. Sander, Phys. Rev. Lett. {\bf 55}, 1315 (1985);
E. Ben-Jacob, P. Garik, T. Mueller and D. Grier, Phys. Rev. A {\bf 38}, 1370
(1988);
E. Ben-Jacob and P. Garik, Physica D {\bf 38}, 16 (1989).

\bibitem{buka1}
\'A. Buka, J. Kertesz and T. Vicsek, Nature {\bf 323}, 424 (1986)
\'A. Buka and P. Palffy-Muhoray, Phys. Rev. A {\bf 36} 1527 (1987);
\'A. Buka, P. Palffy-Muhoray and Z. R\'acz, Phys. Rev. A {\bf 36} 3984 (1987);
\'A. Buka, Physica Scripta {\bf T25}, 114 (1989)

\bibitem{buka2}
\'A. Buka and P. Palffy-Muhoray, J. Phys. France {\bf 49}, 1319 (1988).

\bibitem{buka}
\'A. Buka, in {\it Pattern formation in liquid crystals} (Springer-Verlag,
New York, 1996).

\bibitem{canal}
L. Lam, H. C. Morris, R. F. Shao, S. L. Yang, Z. C. Liang, S. Zheng and H. Liu,
Liquid Crystals {\bf 5}, 1813 (1989);
L. Lam, in {\it Wave Phenomena}, edited by L. Lam and H. C. Morris
(Springer, New York, 1989);
S. L. Yang, Z. C. Liang, R. F. Shao and L. Lam,
in {\it Wave Phenomena}, edited by L. Lam and H. C. Morris
(Springer, New York, 1989).

\bibitem{heine1}
M. Uwaha and Y. Saito, Phys. Rev. A {\bf 40}, 4716 (1989);
E. Brener, H. M\"uller-Krumbhaar and D. Temkin, Europhys. Lett. {\bf 17}, 
535 (1992).

\bibitem{heine2}
T. Ihle and H. M\"uller-Krumbhaar, Phys. Rev. E {\bf 49}, 2972 (1994). 

\bibitem{anisvf}
K. V. McCloud and J. V. Maher, Phys. Rep. {\bf 260}, 139 (1995);
E. Corvera, H. Guo and D. Jasnow, Phys. Rev. E {\bf 52}, 4063 (1995).

\bibitem{bensimon}
D. Bensimon, Phys. Rev. A {\bf 33}, 1302 (1986).

\bibitem{ricard}
R. Gonz\'alez-Cinca, L. Ram\'\i rez-Piscina, J. Casademunt, A.
Hern\'andez-Machado, T. T\'oth-Katona, T. B\"orzs\"onyi and \'A. Buka,
J. Cryst. Growth {\bf 193}, 712 (1998).

\bibitem{pf1}
R. Folch, J. Casademunt, A. Hern\'andez-Machado and L. Ram\'{\i}rez-Piscina,
Phys. Rev. E {\bf 60}, 1724 (1999).

\bibitem{pf2}
R. Folch, J. Casademunt, A. Hern\'andez-Machado and L. Ram\'{\i}rez-Piscina,
Phys. Rev. E {\bf 60}, 1734 (1999).

\bibitem{aref}
G. Tryggvason and H. Aref, J. Fluid Mech., {\bf 136}, 1-30 (1983).

\bibitem{kramer}
T. B\"orzs\"onyi, \'A. Buka and L. Kramer, Phys. Rev. E {\bf 58}, 6236 (1998).

\end{references}
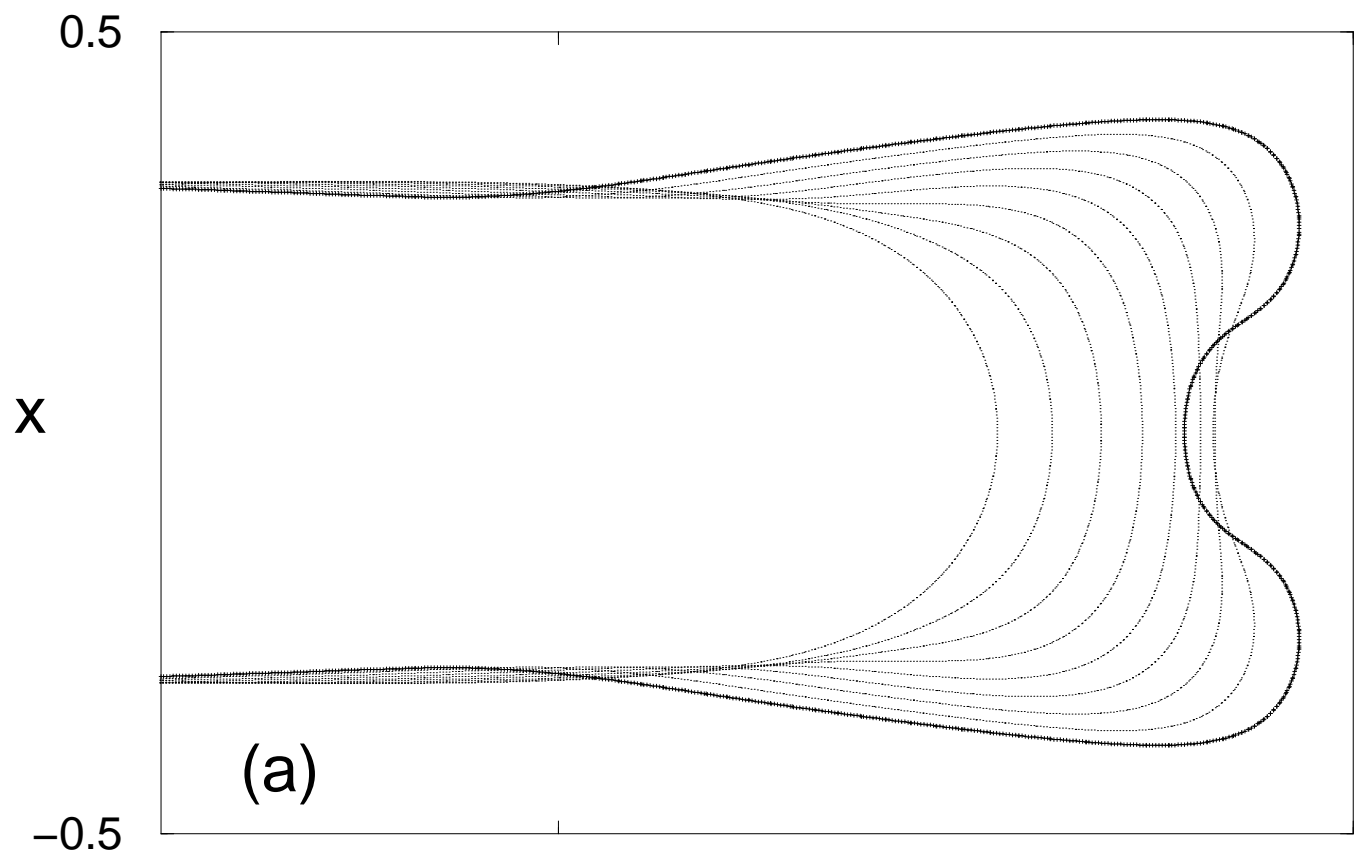
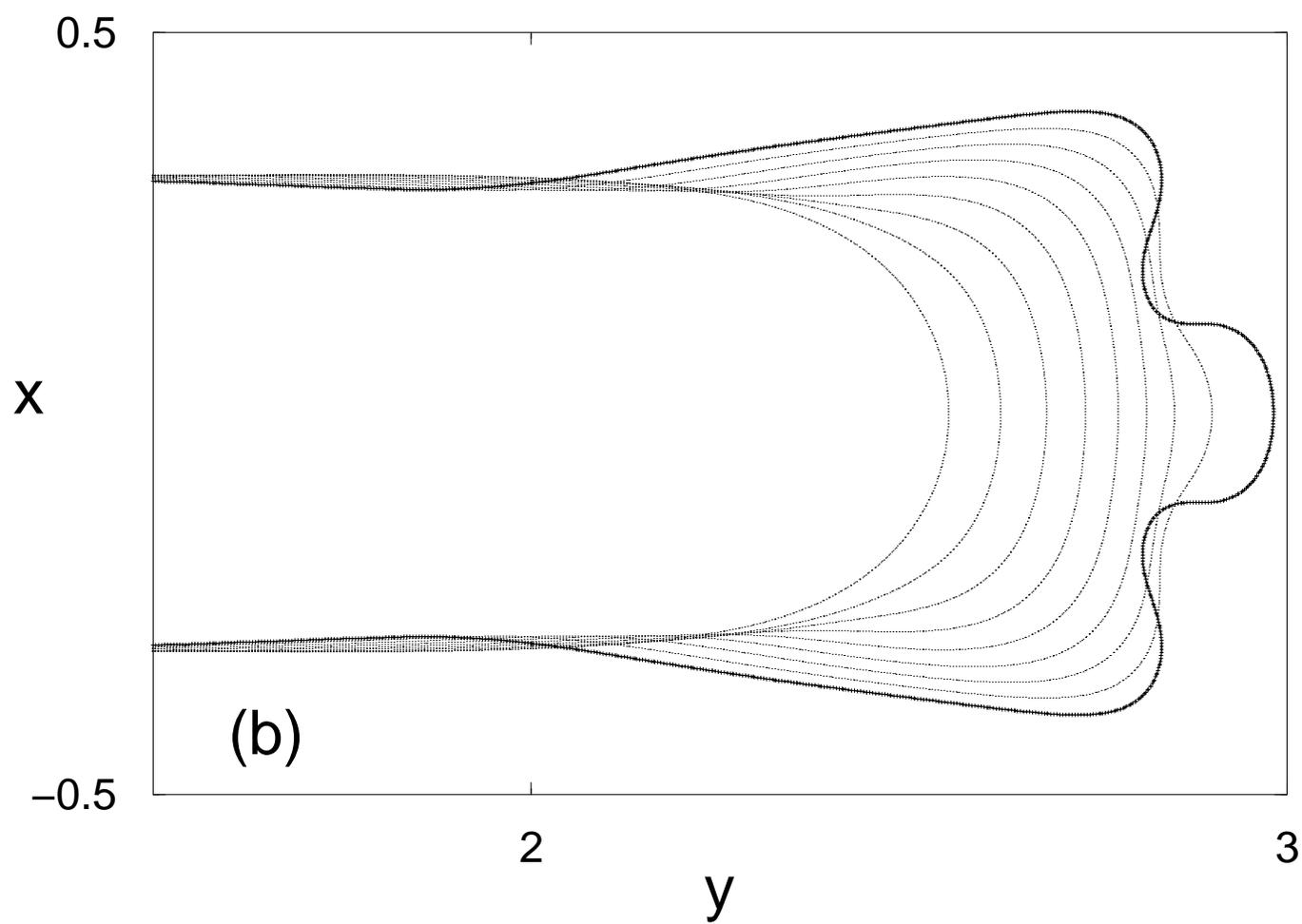
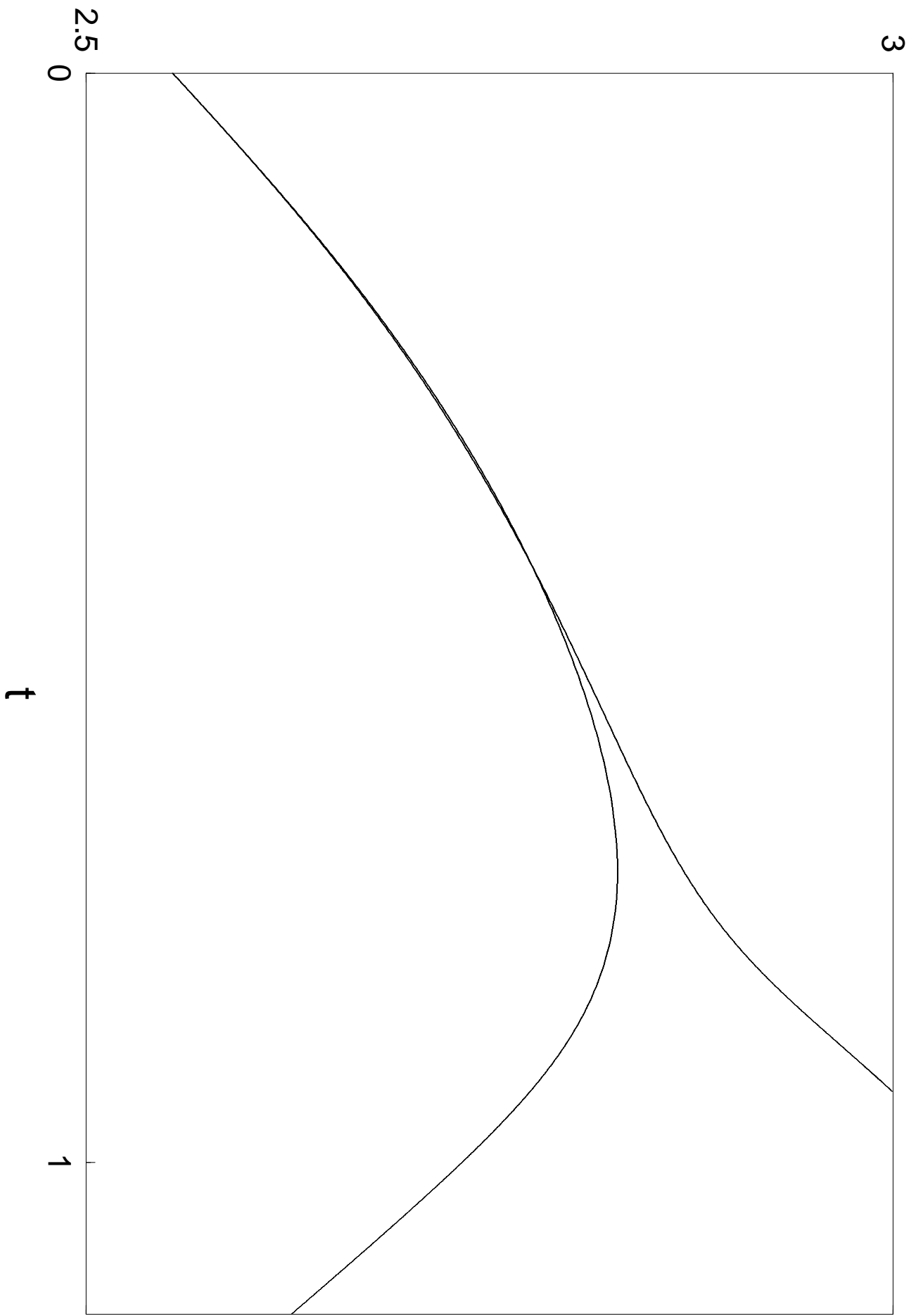
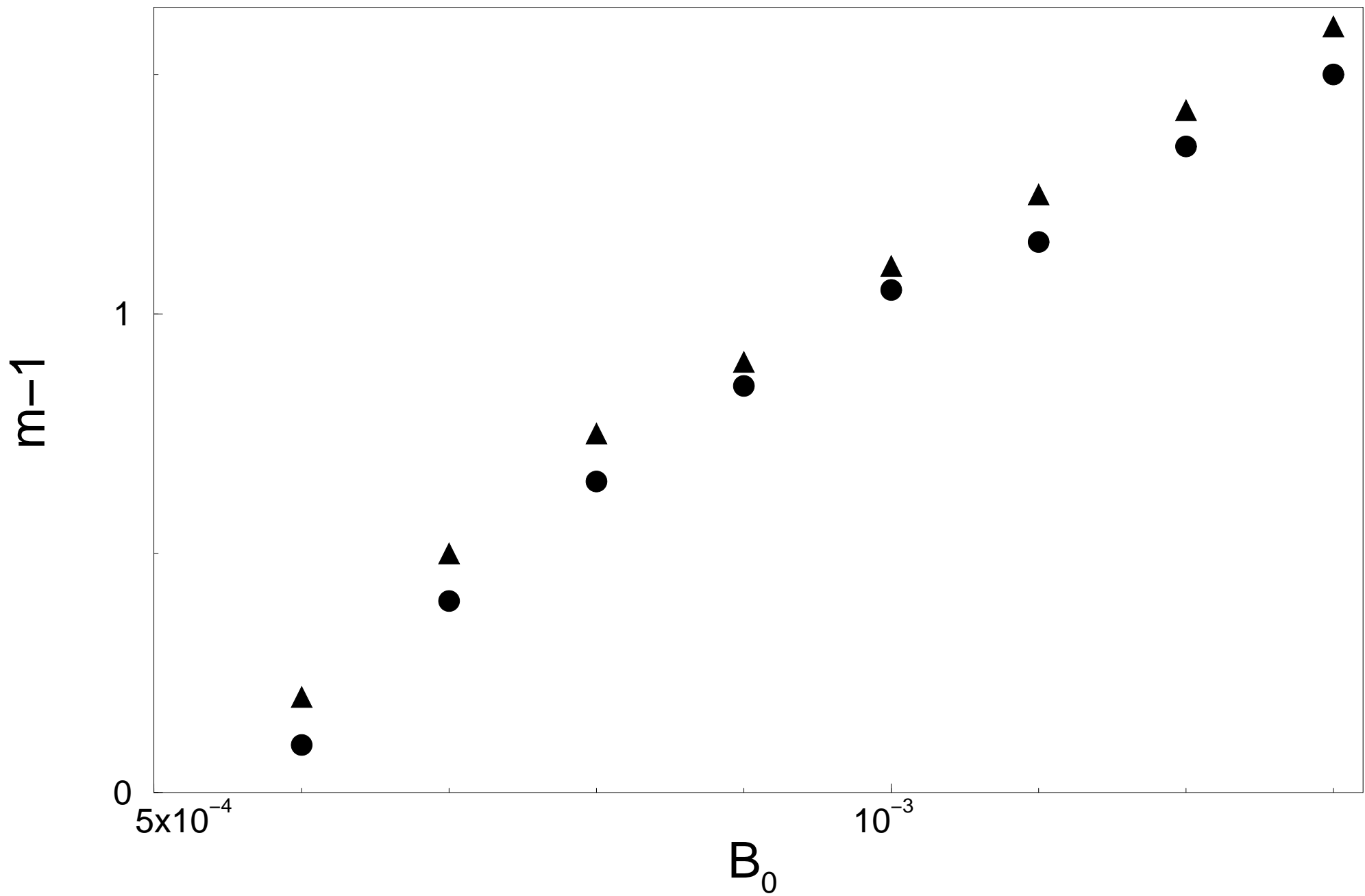
\end{document}